# An improved model of an actively contracting lymphatic vessel composed of several lymphangions: pumping characteristics

C.D. Bertram[1], C. Macaskill[1], J.E. Moore Jr.[2]

[1] School of Mathematics and Statistics, University of Sydney, New South Wales, Australia 2006
[2] Department of Bioengineering, Imperial College, London, England SW7 2AZ

**Abstract**—Using essentially our 2011 numerical model of a multi-lymphangion segment of a collecting lymphatic vessel, but augmented by inclusion of a refractory period and definition of a mid-lymphangion pressure, we explore the effect of several parameters on the form of pump function curves. Pump function is sensitively dependent on the shape of the passive constitutive relation between lymphangion diameter and transmural pressure. Maximum flow-rate increases with the diameter scale applied to the constitutive relation and decreases with the pressure scale. Both maximum flow-rate and maximum pressure difference which can be overcome increase as the excess of lymphangion chain inlet pressure over external pressure is reduced, until inlet pressure is low enough that lymphangion collapse intervenes. The results are discussed in comparison with findings from biological experiments.

## Introduction

The lymphatic vascular system recycles fluid and protein from the body's interstitial spaces to the venous system, in the process exposing the fluid and suspended matter to immune response action at the lymph nodes. A functional subdivision is made between initial lymphatics, which lack muscle cells in their walls and have only sparse internal one-way valves, and collecting lymphatics. Collecting lymphatic vessels have muscular walls and are divided into short segments called lymphangions by frequent regularly spaced one-way valves. These properties enable the autonomous pumping of lymph along a collecting lymphatic vessel. In a previously published paper [1], we established a numerical model of such a vessel consisting of several segments. We showed how the pumping properties were characterized by pump function curves relating mean flow-rate $\bar{Q}$ to the adverse pressure difference faced by the vessel ($\Delta P$). Pumping was improved by increasing the number of lymphangions, by sequential rather than simultaneous lymphangion contraction, and by augmented peak active tension during contraction, and was sensitive to valve specification details and the transmural pressure experienced at the inlet of the lymphangion chain. However, direct comparison of our findings with those from an earlier model [14] was not possible because our model omitted a refractory period between contractions.

We have since rectified this omission, and also improved the model by defining a mid-lymphangion pressure. This latter change was found necessary to enable the model to capture the fluid mechanics of valve prolapse [2]. We examine herein the effects on pump function of that change, and of varying parameters which define the form of the passive relation between lymphangion transmural pressure and diameter. We also show how further increasing the number of lymphangions included in the model, from four to eight, affects the shape of the pump-function curve, Having found that our published explanation [1] of the effects on $\bar{Q}$ and $\Delta P$ of varying the lymphangion-chain inlet pressure relative to external pressure was incomplete, we also provide results which complete the explanation satisfactorily.

## Methods

The model of multiple lymphangions in series used here largely follows the equations and parameter settings of our original model [1]. The equations as modified for inclusion of mid-lymphangion pressure are as follows.

$$\frac{dD_i}{dt} = \frac{2(Q_i - Q_{i+1})}{\pi L D_i} \; ; \quad p_{i1} - p_{im} = \frac{64}{\pi}\frac{\mu L Q_i}{D_i^4} \; ; \quad p_{im} - p_{i2} = \frac{64}{\pi}\frac{\mu L Q_{i+1}}{D_i^4} \quad (1)$$

$$p_{i-1,2} - p_{i1} = R_{Vi}Q_i, \text{ where } R_{Vi} = R_{Vn} + \frac{R_{Vx}}{1 + \exp(-s_o(\Delta p_{Vi} - \Delta p_{oi}))} \text{ and } \Delta p_{Vi} = p_{i-1,2} - p_{i1} \quad (2)$$

$$\Delta p_{tm} = p_{im} - p_e = f_p(D_i) + 2M(t)/D_i, \text{ with } M(t) = M_0[1 - \cos(2\pi f(t - t_{ci}))]/2, \; t_{ci} \leq t \leq 1/f \quad (3)$$

where $i = 1$ to $n$ for valve variables and $i = 1$ to $n-1$ for segment variables, $D(t)$ = diameter, $t$ = time, $Q(t)$ = flow-rate (through a valve), $L$ = lymphangion length, $p_1(t) / p_m(t) / p_2(t)$ = pressure at the upstream end/midpoint/downstream end of a lymphangion, $\mu$ = lymph viscosity, $R_V(\Delta p_V)$ = valve resistance, $R_{Vn}$ = minimum valve resistance, $R_{Vn} + R_{Vx}$ = maximum valve resistance, $\Delta p_V$ = valve pressure drop, $\Delta p_o$ = pressure drop for valve switching, $s_o$ = constant determining slope of $R_V(\Delta p_V)$ at $\Delta p_V = \Delta p_o$, $p_e$ = external pressure, $p_a$ (= $p_{02}$) = inlet reservoir pressure, $p_b$ (= $p_{n1}$) = outlet reservoir pressure, $f_p(D)$ = passive relation between $\Delta p_{tm}$ and $D$, $2M(t)/D_i$ = contribution of active tension, $M_0$ = peak active tension, $f$ = contraction waveform frequency, and $t_{ci}$ = contraction start time, delayed $t_r$ after the end of the previous contraction. Each lymphangion's contractions were delayed 0.5s relative to the one contiguously upstream.

The relation $f_p(D_i)$ between transmural pressure $\Delta p_{tm}$ and diameter $D$ of a passive lymphangion (i.e. no muscle tension) is asymmetrically sigmoidal, reflecting increasing stiffness with increasing distension at positive values (internal exceeds external) of $\Delta p_{tm}$, and progressively diminishing compliance as $\Delta p_{tm}$ becomes increasingly negative and further reduction in cross-sectional area more difficulty. The simple function that we [1] and Jamalian et al. [10] used to model this relation did not change from being collapse-dominated (negative curvature) to stiffening with increasing positive strain (positive curvature) at exactly zero transmural pressure, nor did $\Delta p_{tm} = 0$ coincide with the curve-normalising point $D/D_d = 1$. To rectify this, an alternative equation was devised as

$$\Delta p_{tm} = \frac{4}{15} P_d \left( 12 e^{\frac{D}{D_d} - 1} - 11 - \left(\frac{D_d}{D}\right)^3 \right). \quad (4)$$

The revised relation is illustrated in Figure 1, where it is compared with the original one. Both functions are scaled on the diameter axis by the constant $D_d$, and on the pressure axis by the constant $P_d$.

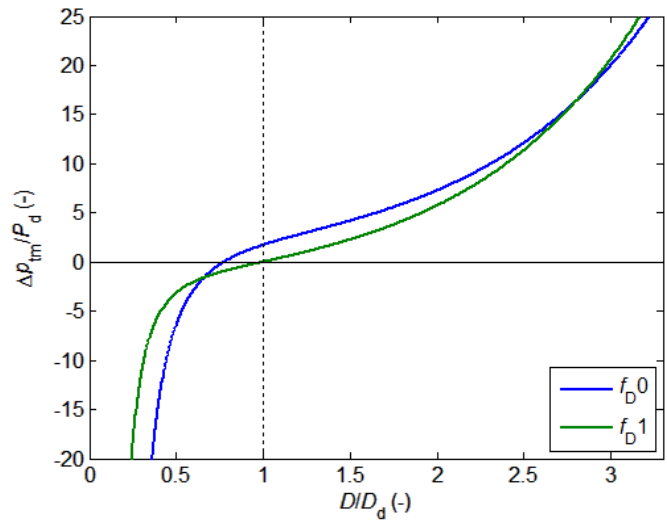

Figure 1  Comparison of the original ($f_D 0$) and revised ($f_D 1$) forms of passive constitutive relation. Lymphangion diameter is normalised by the parameter $D_d$; transmural pressure is normalised by the parameter $P_d$. The slope of $f_D 1$ at $D = D_d$ is $4P_d/D_d$.

The salient parameters are defined in Table 1, which also indicates the default parameter values, and the typical ranges of $\Delta P$ and $\bar{Q}$ achieved in the resulting pump function curves.

| | |
|---|---|
| no. of lymphangions in series, $n-1$ (-) | 4 or 8 |
| number of valves, $n$ (-) | 5 or 9 |
| lymphangion length, $L$ (cm) | 0.3 |
| lymph viscosity, $\mu$ (Poise) | 0.01 |
| valve state-change offset, $\Delta p_o$ (dyn cm$^{-2}$) | $-15$ |
| valve state-change slope const., $s_o$ (cm$^2$/dyn) | 0.2 |
| min. valve resistance $R_{Vn}$ (dyn cm$^{-5}$ s) | 600 |
| valve resistance $R_{Vx}$ increase (dyn cm$^{-5}$ s) | $4.8 \times 10^7$ |
| passive $\Delta p_{tm}$-$D$ relation (see Fig. S1) | $f_D 0$ or $f_D 1$ |
| passive $\Delta p_{tm}$-$D$ diameter scale, $D_d$ (cm) | 0.021 |
| $\Delta p_{tm}$-$D$ pressure scale, $P_d$ (dyn cm$^{-2}$) | 50 |
| peak active tension, $M_0$ (dyn cm$^{-1}$) | 3.6 |
| contraction waveform frequency, $f$ (Hz) | 0.5 |
| refractory period, $t_r$ (s) | 1 |
| inter-lymphangion contraction delay (s) | 0.5 |
| $p_{inlet} - p_{external}$, $\Delta p_{ae}$ (dyn cm$^{-2}$) | 175 |
| typical maximum $\Delta P$ (dyn cm$^{-2}$) | 981 (1 cmH$_2$O) |
| typical maximum $\bar{Q}$ (cm$^3$ s$^{-1}$) | $1.1 \times 10^{-4}$ (0.4 ml/hr) |

Table 1 Default parameter values, and the resulting typical ranges of adverse pressure difference ($\Delta P$) which can be overcome and time-average flow-rate ($\bar{Q}$) which can be generated.

The model was realized in Matlab, using the function ODE23T to solve the differential equations. In this paper we examine the effect of the following factors and parameter variations:
- lymphangions which include or exclude a defined mid-lymphangion pressure,
- varying the shape of the passive $\Delta p_{tm}$-$D$ relation ($f_D 0$ vs. $f_D 1$),
- varying the effective size of lymphangions, via the value of $D_d$,
- varying lymphangion chain stiffness, by varying the value of $P_d$ for all lymphangions,
- varying lymphangion stiffness (via the value of $P_d$) from one lymphangion to the next,
- varying the relative values of inlet pressure $p_a$ and external value $p_e$.

Eight lymphangions are included in the model, and as appropriate we remark on where the increase from four [1] to eight lymphangions caused changes in the pump function, particularly where the effect is on the shape of the resulting curves.

**Results**

The model with a mid-lymphangion pressure defined, which we term the split-lymphangion model, predicts slightly smaller $\bar{Q}$ for any given $\Delta P$, but otherwise does not affect the shape of the pump function curve (Figure 2); all subsequent figures show results with split lymphangions. The overall shape of pump function curves is highly sensitive to the form of the passive $\Delta p_{tm}$-$D$ relation. Used with chains of 3–5 lymphangions [1], the $f_D 0$ form gave curves which mostly or entirely bent away from (were concave to) the origin. Here, with eight lymphangions, the $f_D 0$ form yields curves (shown in blue) that display reverse curvature, bending toward the origin at low $\bar{Q}$ and away from the origin at high $\bar{Q}$. In contrast, the $f_D 1$ form leads to curves largely bending toward the origin. At the cost of some reduction in $\Delta P$ overcome at modest values of $\bar{Q}$, the $f_D 1$ form almost doubles the maximum $\bar{Q}$ attained when $\Delta P = 0$.

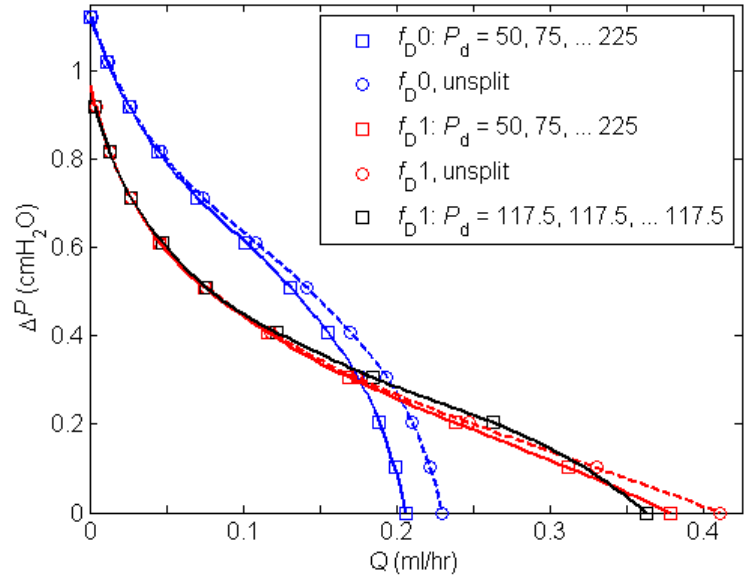

Figure 2   Pump function curves for eight lymphangions in series.  Solid curves denote split lymphangions; dashed curves denote unsplit ones.  The blue curves use the $f_D0$ passive $\Delta p_{tm}$-$D$ relation; the red and black curves use the $f_D1$ form.  Blue and red curves have the parameter $P_d$ (dyn cm$^{-2}$) increasing along the chain as indicated in the legend; the black curve compares an equivalent model having the same $P_d$ value for all lymphangions.  Other parameters as in Table 1.

For the models providing the results illustrated by blue and red curves in Fig. 2, each successive lymphangion in the direction of flow was given a higher value of the pressure scaling parameter $P_d$. Multi-lymphangion lymphatic vessel segments that are pumping actively between a low-pressure inlet reservoir and a higher-pressure outlet reservoir will in general experience higher time-averaged $\Delta p_{tm}$ in each successive lymphangion.  With $P_d$ increasing going downstream, each successive lymphangion is stiffer (in both distension and collapse), and therefore better able to withstand increasing transmural pressure.  Such variations have not yet been detected experimentally.  The effect of a $P_d$-distribution on the shape of the pump function curve was otherwise minor; cf. the black and solid red curves in Fig. 2.

As shown by Fig. 1, the parameter $D_d$ essentially sets the diametric size of the lymphangions, and therefore their hydraulic conductance; all other things being equal, this determines the maximum $\bar{Q}$ which they can convey.  However, by the law of Laplace, an increase in lymphangion diameter reduces the $\Delta p_{tm}$ contribution of a given level of active tension (Appendix 2), and therefore reduces the maximum adverse pressure difference $\Delta P$ that the chain of lymphangions can overcome.  These effects are both clearly shown in Figure 3, where varying $D_d$ by ±38% changed the maximum $\Delta P$ by some – or +22% respectively; the 38% $D_d$-increase led to a 39% higher maximum $\bar{Q}$.  During contraction, some lymphangions visited the collapsed state, as indicated in the figure.

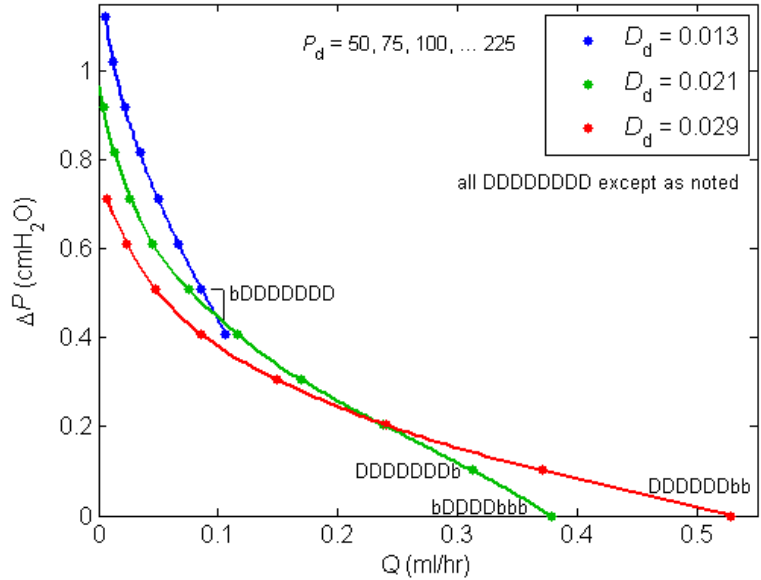

Figure 3   Pump function curves for an 8-lymphangion model using the $f_D1$ relation, with $\Delta p_{ae}$ = 175 dyn cm$^{-2}$, $M_0$ = 3.6 dyn cm$^{-1}$, $P_d$ values (dyn cm$^{-2}$) increasing along the chain, and three different values (cm) of $D_d$. For each lymphangion it is noted whether during a cycle $D(t)$ remained always above $D_d$ (distended, D) or whether it also visited the collapse region (both, b); thus DDDDDDDb means that only $D_8(t)$ visited both sides of $D_d$ (lymphangion number increases in the direction of $\bar{Q}$).

The predominant effect of increasing the pressure-scaling parameter $P_d$ is to reduce the diastolic filling of lymphangions for a given level of diastolic transmural pressure (set by the pressure difference $\Delta p_{ae} = p_a - p_e$), rather than to increase the levels of $\Delta P$ that the chain can overcome. As $P_d$ increases, the stiffer lymphangions convey less fluid at small and moderate levels of $\Delta P$, as shown in Figure 4.

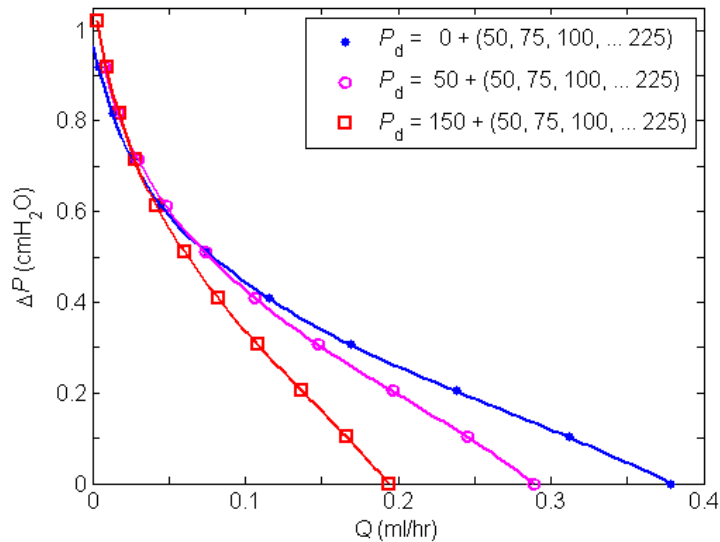

Figure 4   Pump function curves for an 8-lymphangion model using $f_D1$ as before, with symbols corresponding to values of $P_d$ (dyn cm$^{-2}$) as shown in the legend.

The effect of varying the overall level of transmural pressure for the chain, by varying $\Delta p_{ae}$, was explored in fig. 10 of Bertram et al. [1]. There, using the $f_D0$ passive $\Delta p_{tm}$-$D$ relation, the effect of decreasing $\Delta p_{ae}$ from 175 to 87.5 to 0 dyn cm$^{-2}$ was uniformly to decrease maximum $\bar{Q}$, but also to increase maximum $\Delta P$. The issue is here revisited for the $f_D1$ relation (Figure 5). It is now apparent that the previously-seen trend for $\bar{Q}$ was only part of the story. Here, lower values of $\Delta p_{ae}$ generally lead to curves with a <u>higher</u> maximum $\bar{Q}$. But when $\Delta p_{ae}$ and $\Delta P$ are both sufficiently low that most lymphangions are collapsed in diastole ($\Delta p_{ae}$ = 87.5 dyn cm$^{-2}$), the pump function curve folds back on itself in such a way as to give a reduced maximum $\bar{Q}$. Curves for lower $\Delta p_{ae}$ than this would fold back at a lower $\bar{Q}$ again. The same trend as before is seen for maximum $\Delta P$: lower $\Delta p_{ae}$ leads to increased maximum $\Delta P$.

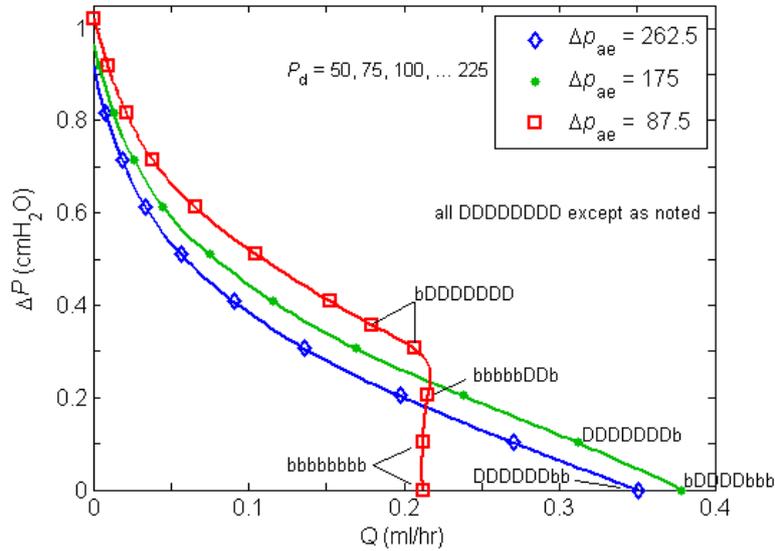

Figure 5   Pump function curves for the same $f_D1$ model as above, at three levels of inlet transmural pressure $\Delta p_{ae}$ (dyn cm$^{-2}$) as indicated in the legend; all other parameters as for Fig. 3. The bD codes are explained in the caption to Fig. 3.

**Discussion**

In the absence of a relation between muscle length and active tension, i.e. with active tension a function only of time, the overall shape of the pump function curve is sensitively dependent on the passive relation between transmural pressure and diameter. If the curvature of the $\Delta p_{tm}$-$D$ relation at $D = D_d$ and at low $\Delta p_{tm} > 0$ is negative, i.e. d$^2\Delta p_{tm}$/d$D^2 < 0$, as in the $f_D0$ relation, the pump function curve bends outward from the origin. This was seen as a non-physical intrusion of the negative curvature associated with vessel collapse into the region where $\Delta p_{tm} > 0$ and positive curvature (d$^2\Delta p_{tm}$/d$D^2 > 0$) associated with stiffening with increasing stretch was expected to dominate. The $f_D1$ relation was designed specifically to remove this feature of the passive $\Delta p_{tm}$-$D$ relation. Using the $f_D1$ relation, pump function curves bend inward toward the origin (Figs. 2–5). The only exception occurs when $\Delta p_{ae}$ is sufficiently low that contraction brings about collapse of the lymphangion chain, limiting the flow-rate (Fig. 5).

Other parameters mainly affect not so much the shape of the pump function curves as the values of $\Delta P$ and $\bar{Q}$ which are reached. If high $\Delta P$ is to be overcome, many lymphangions will be needed in a series chain [1]. If high $\bar{Q}$ is required, then large-diameter lymphangions are needed, i.e. increased values of $D_d$ ([10] and Fig. 4); lymphangions with lower passive stiffness (reduced $P_d$) and therefore greater diastolic compliance also help to achieve high $\bar{Q}$ (Fig. 4). Increased peak active tension increases both maximal $\Delta P$ and maximal $\bar{Q}$, proportionally in the case of $\Delta P$, rather less so for $\bar{Q}$ [1]; increased transmural pressure (in the form of $\Delta p_{ae}$) reduces maximal $\Delta P$ and $\bar{Q}$ (Fig. 5).

**Comparison with literature**

Several investigators have measured the variation of time-averaged lymph flow-rate with changes in the downstream pressure presented to a lymphatic vessel, but the resulting pump function curves are often incomplete, stopping short of either the maximal-$\bar{Q}$ limit or the maximal-$\Delta P$ limit or both. Drake et al. [4-6], cannulating the outlet of entire lymphatic beds, could not measure the inlet pressure to their networks, so the pressure axis on their pump function curves is not convertible to pressure difference. Such curves represent networks with parallel vessels, collaterals, and small non-muscular vessels all included, which are not described by our model. Furthermore, the pumping they measured was mainly not active (intrinsic), but the result of external compression.

Figure 6a is a redrawing of data from Venugopal et al. [13], showing partial pump function curves for four actively pumping vessels (all at $\Delta p_{ae}$ = 5 mmHg[1]). A common trend is hard to discern. Figure 6b shows data of $\bar{Q}$ vs. outlet pressure measured by Eisenhoffer et al. [7] for one bovine mesenteric lymphatic vessel at various transmural pressures. Their 'transmural pressure' approximates $\Delta p_{ae}$ except insofar as there is an inlet catheter pressure drop. Their 'outlet pressure' approximates $p_b - p_e$; the curves therefore begin at values which exceed the transmural pressure, in order to provide $\Delta P > 0$. The pressure axis would be $\Delta P$ if the transmural pressure were subtracted from the outlet pressure. They reported that $\bar{Q}$ was consistently nonlinearly related to the outlet pressure, with $\bar{Q}$ being maintained roughly constant until outlet pressure was high. This report is consistent with the behaviour of the model with length/active-tension relation $f_M 2$. These curves also display trends entirely in agreement with those of Fig. 5; maximum $\Delta P$ increases as transmural pressure decreases, and lower transmural pressures lead to curves further from the origin, except that the lowest transmural pressure leads to $\bar{Q}$ being limited to modest values by lymphangion collapse during part of the contraction cycle. However, in any such comparison it should be remembered that our model is currently geared toward the emulation of physiological data for small rat mesenteric lymphatics, which do not necessarily behave the same way as these much larger lymphatics from other species. Also, our models maintained constant contraction frequency. The frequency of lymphangion contraction in the experiments is rarely recorded; if one assumes more frequent contractions at higher outflow pressures [11], then conversion to constant frequency would tend to increase curve convexity to the origin.

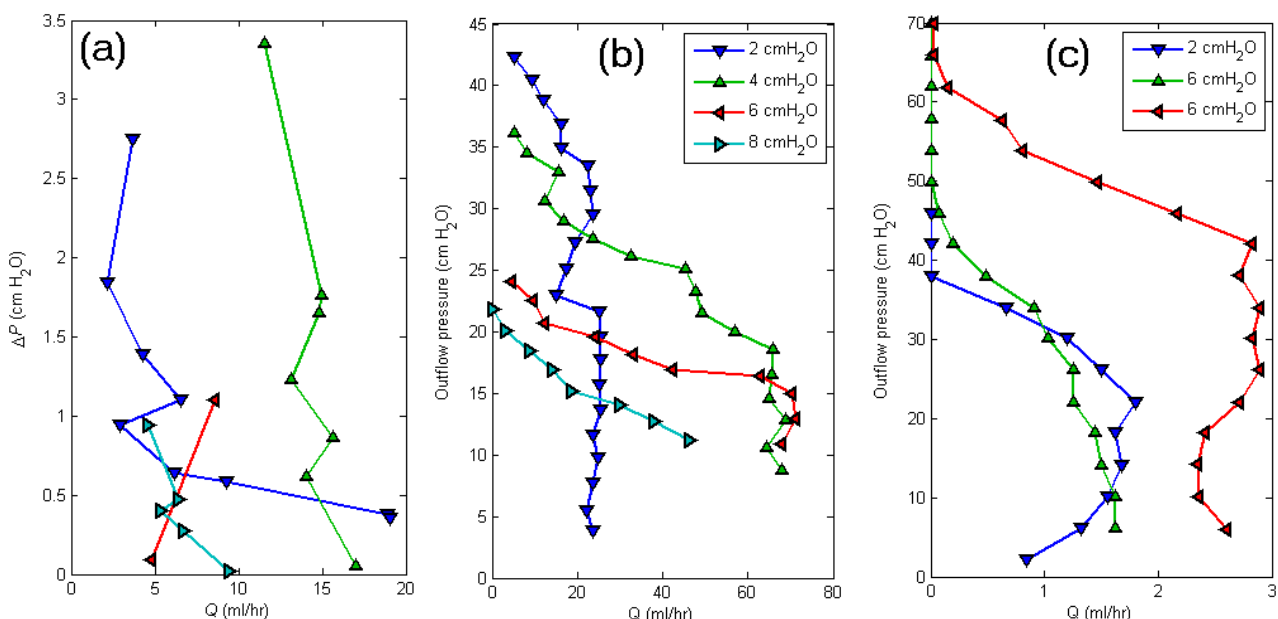

Figure 6  (a) Pump function for bovine mesenteric lymphatics *in vitro* [13]. (b) Pump function, one bovine mesenteric vessel [7]. Transmural pressure as indicated in the legend. (c) Pump function, three ovine prenodal popliteal vessels [9]. Inflow pressure as indicated in the legend.

---

[1] Possibly minus 1 cmH$_2$O of external hydrostatic pressure; the text [14] is somewhat ambiguous. Quick et al.[12] stated that their reported inlet and outlet pressures were transmural pressures.

Data from similar experiments by Eisenhoffer et al. [9] on isolated sheep popliteal lymphatic vessels (Figure 6c) show a much smaller $\bar{Q}$ range, but their pump function also predominantly bends away from the origin. It is tempting to ascribe the diminishing $\bar{Q}$ at low outflow pressure and 2 cmH$_2$O inflow pressure to the valves allowing backflow [3]. Curves of outflow pressure vs. $\bar{Q}$ that bend away from the origin were also found by McGeown et al. [11] for ovine popliteal efferent and distal hind-limb afferent lymphatics outlet-cannulated *in situ*. In those experiments, the frequency of contractions generally increased with outflow pressure.

Figure 7 shows a pump function curve for lymphatic 'systems' each consisting of five segments connected in series, with each segment consisting of 4–9 lymphangions [8]. Given the scatter of the data, a variety of underlying shapes is possible here. Again it is tempting to see the reverse curvature here as echoing that seen in our simulations here of eight lymphangions (Fig. 2).

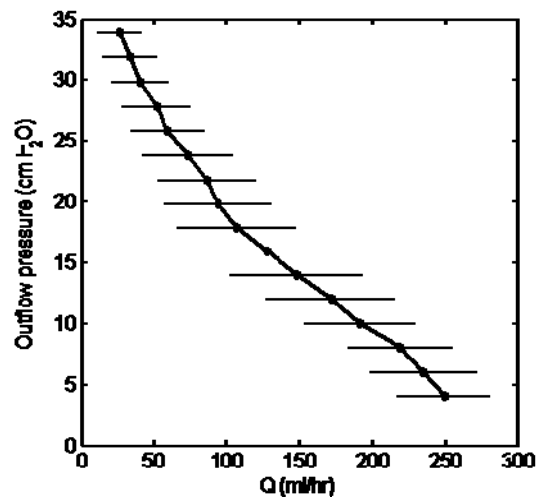

Figure 7  Pump function for six 'systems' of five bovine mesenteric lymphatic segments in series [8]. Inflow pressure was 4 cmH$_2$O. Bars ± s.e.m.

There is an experimental difficulty in achieving a pump function curve for an individual vessel, in that the cannulas will necessarily add their own fixed resistance to the flow circuit. Drake et al. [5] were aware of this, and allowed for the measured resistance of their cannulating pipette. An alternative technique would be to measure the pressure beyond each pipette tip using the servo-null method. This would have the advantage of capturing the time variation of that pressure as well as the mean value, allowing its comparison with simulation. However, even this does not overcome the more significant problem that arises if cannula resistance is large, such that the mean pressure drop between the pressure reservoir and the cannula tip is large relative to the mean pressure difference across the lymphangion chain. Under these circumstances, the pumping by the lymphangions will not control $\bar{Q}$. The unnatural situation then exists whereby a pump is being characterised under conditions where it does not set the flow-rate. Data on $\Delta P$ and $\bar{Q}$ can still be obtained, but it is open to question whether the resulting pump function curve corresponds to that which would have been obtained if ideal constant-pressure sources/sinks had been applied to the chain itself. It is possible to simulate an isolated lymphatic vessel that is subject to such inlet and outlet resistances, but the best solution, both experimentally and numerically, is to ensure that these external resistors are as far as possible negligible.

Venugopal et al. [14] reported that their inlet and outlet resistances were 192 and 207 dyn s/cm$^5$ respectively (the seconds were omitted but this can safely be assumed to be a typographical error); however it is unclear whether this included all the tubing or only that between the tip and the site of pressure measurement[2]. Fig. S6a shows $\bar{Q}$ up to 19 ml/hr; thus the inlet and outlet pressure drops each would have reached ~1 dyn/cm$^2$, far below the order of $\Delta P$ (0–3.4 cmH$_2$O or 0–3,335 dyn/cm$^2$). Similarly, Eisenhoffer et al. reported inlet and outlet resistances of 56 and 72 dyn s/cm$^5$ when working with bovine vessels [7], and 138 and 148 dyn s/cm$^5$ with ovine vessels [9]. This check is always needed, particularly when micropipettes are in use.

Comparing each of the curves in Figs. 6b and 6c with a straight line between the extreme data points, it is seen that three of the four curves in Fig. 6b bend slightly away from the origin, and two of the three curves in Fig. 6c bend markedly away. The experimental evidence thus offers some support for simulated pump function curves that bend strongly away from the origin. Pumps with this characteristic produce relatively high maximum power within the limits sets by the maximum $\bar{Q}$ they can generate and the maximum $\Delta P$ they can overcome. In a subsequent paper, we will describe further model developments that tend to lead to pump function curves having this feature.

**Conclusions**

The transport capacity of multi-lymphangion contracting segments of lymphatic vessel can be characterised by pump function curves. The shape of such pump function curves, which is linked to their power production, depends sensitively on the form of the constitutive relation describing how lymphangion diameter varies with transmural pressure. Other parameters, such as the constitutive-relation scaling constants and the extent to which inlet pressure exceeds external pressure, influence more the maximum flow-rate which can be generated and the maximum adverse pressure difference which can be dealt with.

**Acknowledgements**

JEM gratefully acknowledges support from U.S. National Institutes of Health grant R01-HL-094269, the Royal Academy of Engineering and a Royal Society-Wolfson Research Merit Award. All authors acknowledge support from U.S. National Institutes of Health grant U01-HL-123420.

---

[2] Quick et al. [12] show a lymphatic external diameter vs. time trace with mean diameter ~2 mm stemming from related experiments. If it be assumed that the <u>internal</u> diameter of the inlet and outlet tubing was also 2 mm, and that perfusate viscosity was 1 cP, then Poiseuille gives a total tubing length at each end of < 1 cm.